\documentclass[11pt]{article}
\usepackage{graphicx}
\usepackage{multicol}
\title{\bf On the Equivalence Principle and a Unified Metric Description of Gravitation and Electromagnetism}
\author{  Murat \"Ozer\\
Department of Physics,\\
Faculty of Arts and Sciences,\\
Y{\i}ld{\i}z Technical University,\\
34220 Esenler, Istanbul, Turkey\\
E-mail: mhozer@yildiz.edu.tr\\
}
\date{ }
\begin{document}
\maketitle

\begin{abstract}
We first investigate the form the General Relativity theory would have
taken had the gravitational mass and the inertial mass of material objects
been different. We then extend this analysis to electromagnetism and
postulate an equivalence principle for the electromagnetic field. We argue
that to each particle with a different electric charge-to-mass ratio in
superimposed gravitational and electromagnetic fields there corresponds a
spacetime manifold whose metric tensor $g_{\mu\nu}$ describes the dynamical
actions of gravitation and electromagnetism. The electric field outside a charged
sphere asserts itself independently rather than contributing to the gravitational
field. The contribution of the electric field to the spacetime metric outside the
charged sphere is shown to be similar to the gravitational one in
the Schwartzschild metric but with a charge-to-mass ratio dependence of the test
particle instead of the Reissner - Nordstr\"om metric, resulting in a unified
description of gravitation and electromagnetism.  We point out that there are existing
experiments whose results can be explained by the equivalence principle for the 
electromagnetic field presented here. Additional experimental predictions of the theory are
 mentioned.
\end{abstract}

\section{Introduction
 \label{sec:intro}}
The possibility of exhibiting gravitation and electromagnetism in a unified
geometrical representation has been pursued by many mathematicians and
physicists. The first one to seek for a unified explanation of gravitation
and electromagnetism was Riemann (see Ref.\cite{mis}). This endeavor has
really started as a full-fledged research area soon after the advent of
Einstein's general relativity theory  \cite{ein1}. The gauge-invariant unified
theory of Weyl was based on a generalization of Riemannian geometry
\cite{ber,wey}. A generalization of  Weyl's theory was put forward by
Eddington \cite{edd}. These unsuccessful
attempts were followed by Kaluza who sought to include the electromagnetic
field by increasing the number of components of the metric tensor by
changing the number of dimensions to five \cite{kal}, whose work was later
revived and extended by Klein \cite{kle}. Generalizations of Kaluza's
theory were attempted by Einstein and coworkers \cite{ein2,ein3,ein4}.
Another line of approach that
produced the same field equations as in Kaluza's theory was that of
projective field theories \cite{veb,pau}. Also worth mentioning is the
work of Einstein based on Riemannian metrics and distant parallelism
\cite{ein5}. Since the electromagnetic field is described by a second rank
antisymmetric tensor, the idea of employing a nonsymmetric metric tensor
$g_{\mu\nu}$ whose antisymmetric part is to be associated with
electromagnetism was exercised too \cite{schr}.

All these attempts to obtain a unified theory of gravitation and
electromagnetism in a four dimensional Riemannian spacetime have been
baffled hitherto. What lies at the root of this bafflement is the fact
that electrically charged particles do not possess a universal
charge-to-mass ratio. It is often stated that a geometric theory of
electromagnetism could have been achieved had this ratio been the same for
all particles.
One immediate consequence of  such a hypothetical unified theory
through a symmetric $g_{\mu\nu}$ in four dimensions would have been having
to relax the interpretation of the $g_{\mu\nu}$ as the gravitational
field since then the components of which would have corresponded to
superimposed gravitational and electromagnetic potentials.

Having failed in these attempts, instead of yielding to a complete failure,
it is our opinion that, be it not universal, a restricted or
specific geometrization of electromagnetism should be sought. The way to
achieve this seems to consider the motion of each charged test particle
one by one in a given electromagnetic field and geometrize each case
seperately. The resulting geometrization, of course, will not be of the
universal type as in gravitation. We hope to present in this work the
grounds for reasons to believe that this endeavor may be full of
experimentally testable surprizes.

To this end, we shall first postulate an equivalence principle for the
electromagnetic field by way of examples and thereby  conclude that the
general relativity theory, after a  correction, is not only the
theory of gravitation but also a unified theory of gravitation and
electromagnetism. In order to reach  this conclusion we ought to
emancipate ourselves from two conceptual obstacles. First, that the
equality of the gravitational mass and the inertial mass is indispensable
for the formulation of general relativity (hereafter GR). Second, that the
metrical field $g_{\mu\nu}$ represents the gravitational potentials only.
Of course, it is indeed a remarkable fact of nature that the gravitational
and inertial masses associated with all material objects are equal to a
great accuracy \cite{eot,rol,bra}.
As a result of this equality a given gravitational field imparts the same
acceleration to all particles at a given spacetime point. Einstein
generalized the experimental results on the equality of these two masses to
the (weak) Equivalence principle \cite{ein6}, that {\em a uniform
gravitational field and a uniformly accelerating frame are locally
equivalent}, or stated slightly differently {\em gravitational and
inertial forces are locally completely equivalent.}

This paper is organized as follows. In Sec. \ref{sec:hypo} we consider
a hypothetical world where the gravitational and inertial masses of material
objects are different. We write down the Einstein field equations
in such a world. In Sec. \ref{sec:newton} we treat Newtonian electromagnetism
in the language of curved spacetime. The energy of a test charge in the viscinity
of a charged sphere is considered to obtain the $g_{00}$ component of the resulting
spacetime metric in Sec. \ref{sec:sphere}. In Sec. \ref{sec:action} we consider the 
action integral for a charged particle in superimposed gravitational and electromagnetic
fields and obtain the same $g_{00}$ once more. In Sec. \ref{sec:elevator} we propose
an elevator thought experiment for charged particles in an electromagnetic
field. In Sec. \ref{sec:gravitational} we propose the modified field
equations for a distribution of mass and electric charge. The line
element for a spherically symmetric distribution of matter and charge
is presented in Sec. \ref{sec:line}.
Concluding remarks are presented in Sec. \ref{sec:concluding}.

\section{Hypothetical General Relativity with $m_g\ne m_i$
\label{sec:hypo}}
The question of the gravitational mass $m_g$ of a body not being equal to
its inertial mass $m_i$ due to the possibility that the gravitational self
energy of the body contributes unequally to $m_g$ and $m_i$ was addressed
in a series of papers by Nordtvedt \cite{nor}. He argues that if it is
assumed
\begin{equation} 
\frac{m_g}{m_i}=1+\eta\frac{G}{c^2}\int \rho(\vec x)\rho(\vec x')
\frac{d^3xd^3x'}{\mid \vec x-\vec x'\mid}/\int \rho(\vec x)d^3x,
\label{eqn:nord}
\end{equation}
where $\eta$ is a dimensionless constant of order of magnitude 1, $G$ is the
gravitational constant, $c$ is the speed of light, and $\rho(\vec x)$ is the
mass density of the body, then for the bodies used in the experiments of
Refs. \cite{eot,rol,bra}, the correction term in Eq.(\ref{eqn:nord})
is of order $10^{-25}$, thereby not contradicting these experiments. We
would like to argue in the following, by way of a thought experiment, that
the basic structure of GR would have remained intact had the gravitational
mass been not equal to the inertial mass. \\

Consider an elevator cabin falling
freely in a given gravitational field $\vec g$. Let there be test particles
inside the cabin with different $m_g/m_i$ ratios
\footnote{Notice that the gravitational mass $m_g$ here is actually the passive
gravitational mass, which is the mass acted upon by a gravitational field. 
Because of their equality in Newtonian mechanics, we do
not distinguish in this work between the passive and active gravitational masses, 
the latter being the mass that gives rise to a gravitational field, and use the term
 gravitational mass only}. Let the elevator, 
an observer in it, and {\em one} of the particles have the same  $m_g/m_i$ ratio. As
the elevator falls, let the observer drop the test particles simultaneously
from rest. He/she will then see the particles strike the floor or the
ceiling of the elevator one by one according to their acceleration relative
to the elevator (or the observer)
\begin{equation} 
a_{rel}=\left(\frac{m_g}{m_i}-\frac{M_g}{M_i}\right)g,
\label{eqn:elev1}
\end{equation}
where $M$ is the mass of the elevator. But the test particle having the
same ratio as the elevator will float motionlessly. What has happened is
that the gravitational force on this particle has been cancelled by the
inertial force on it due to the downward acceleration of the elevator.
We, therefore, conclude that this freely falling elevator is an inertial
frame only for this particular particle, but not for the others. Stated
equivalently, had the $m_g/m_i$ ratio of the particles been different in a
hypothetical world there would have been locally inertial but nonidentical
frames unique to each particle or particles with the same $m_g/m_i$ ratio.
This is in contrast to what happens when gravitational and inertial masses
are equal in which case the local inertial frames in the neighborhood of
each particle are identical and the particles move freely in the same
geometry. But, since a given point may contain only one particle at a given
time, and each particle obeys its own equation of motion, there is no
reason why particles could not have travelled in their own geometry had
their $m_g$ been different from their $m_i$. Each particle, then, would
have followed its own geodesic according to
\begin{equation} 
\frac{d^2x^{\mu}}{d\lambda^2}+\Gamma^{\mu}_{\alpha\beta}
\frac{dx^{\alpha}}{d\lambda}\frac{dx^{\beta}}{d\lambda}=0,
\label{eqn:geo}
\end{equation}
where the Christoffel symbols (connection coefficients)
$\Gamma^{\mu}_{\alpha\beta}$ would have depended
on the $m_g/m_i$ ratio of the test particle and $\lambda$ is an affine
parameter, such as the proper time $\tau$ or the proper length $s$, of the
geodesic. The  equivalence principle
then would have been {\em it is impossible to distinguish the fictitious
inertial forces from the real gravitational forces in a local region
containing a single particle or particles with the same $m_g/m_i$ ratio.}
This we shall call the {\em single-particle equivalence principle.} What
would have happened to the Einstein field equations in such a hypothetical
world? By considering the Newtonian limit it can be seen that the field
equations would have taken the form
\begin{equation} 
R^{\mu\nu}-\frac{1}{2}g^{\mu\nu}R=\frac{8\pi G}{c^4}T'^{\mu\nu},
\label{eqn:ein}
\end{equation}
where
\begin{equation} 
T'^{\mu\nu}=\frac{m_g}{m_i}T^{\mu\nu},
\label{eqn:tens}
\end{equation}
with $T^{\mu\nu}$ being the energy-momentum tensor of a distribution of
matter or other forms of energy
\footnote{Incidentially, Eqs. (\ref{eqn:ein}) and (\ref{eqn:tens})
indicate that $m_g/m_i=1$ is imposed in Einstein's GR.}.
Hence the solutions of Eq.(\ref{eqn:ein}) would have
involved $m_g/m_i$ of the test particle. For example, the Schwartzschild
exterior solution \cite{schw}  for a static spherical distribution of mass $M_g$
would have been
\begin{eqnarray} 
ds^2=-\left(1-2\frac{m_g}{m_i}\frac{GM_g}{c^2r}\right)c^2dt^2+
\left(1-2\frac{m_g}{m_i}\frac{GM_g}{c^2r}\right)^{-1}dr^2+
r^2d\theta^2+r^2sin^2\theta d\phi^2.
\label{eqn:schw}
\end{eqnarray}
Note, in such a world, that test particles with different $m_g/m_i$'s would
have had different spacetime geometries when they are in the same
gravitational field.

\section{Newtonian Electromagnetism in the Language of Curved Spacetime
\label{sec:newton}}
Having presented what would have happened to GR in a hypothetical world
where $m_g\neq m_i$, we can immediately draw a parallelism with
electromagnetism in which the {\em field charge} is the electric charge $q$
of a particle instead of the gravitational mass $m_g$. To gain further
insight into our problem and to convince ourselves that we are on the right
path, let us translate Newtonian gravity in the language of curved
spacetime (a la Cartan) \cite{tra,see} into Newtonian electromagnetism in the
language of curved spacetime. The trajectory of a charged particle in an electromagnetic
field subject to the force
\begin{equation} 
\vec F=q\vec E+q\vec v\times\vec B,
\end{equation}
is given in Newtonian electromagnetism by (i=1,2,3 and summation over repeated
indices is implied.)
\footnote{From now on we denote the inertial mass $m_i$ by $m$,
whenever there is no confusion.}
\begin{equation} 
\frac{d^2x^i}{dt^2}+\frac{q}{m}\left(\frac{\partial\Phi_E}{\partial x^i}+
\frac{1}{c}\frac{\partial A^i}{\partial t}\right)+
\frac{q}{m}\epsilon^{ijk}\epsilon^{kln}\frac{dx^j}{dt}\frac{\partial A^n}
{\partial x^l}=0,
\label{eqn:newt}
\end{equation}
where $t$ is the coordinate time, $\Phi_E$ is the electric  potential, and
$\vec A$ is the  vector potential. This can be written in curved spacetime as,
\footnote{After the use of $\epsilon_{ijk}\epsilon_{kln}=\delta_{il}\delta_
{jn}-\delta_{in}\delta_{jl}$.}
\begin{equation} 
\frac{d^2t}{d\lambda^2}=0,\hspace{0.5cm} \frac{d^2x^i}{d\lambda^2}
+\frac{q}{m}\left(\frac{\partial\Phi_E}{\partial x^i}
+\frac{1}{c}\frac{\partial A^i}{\partial t}-\frac{dx^n}{dt}
\frac{\partial A^n}{\partial x^i}+\frac{dx^n}{dt}\frac{\partial A^i}{
\partial x^n}\right)\left(\frac{dt}{d\lambda}\right)^2=0,
\label{eqn:curv}
\end{equation}
where the geodesic parameter $\lambda=at+b$, $a$ and $b$ being arbitrary
constants. By comparing Eq. (\ref{eqn:curv}) with the geodesic equation
(\ref{eqn:geo})
the nonzero connection coefficients are read off as
\begin{equation} 
\Gamma^i_{00}=\frac{q}{mc^2}\left(\frac{\partial\Phi_E}{\partial x^i}
+\frac{1}{c}\frac{\partial A^i}{\partial t}
-\frac{dx^n}{dt}\frac{\partial A^n}{\partial x^i}+\frac{dx^n}{dt}\frac{
\partial A^i}{\partial x^n}\right).
\label{eqn:gam00}
\end{equation}
By inserting these in the Riemann tensor
\begin{equation} 
R^{\mu}_{\nu\alpha\beta}=\frac{\partial\Gamma^{\mu}_{\nu\beta}}
{\partial x^{\alpha}}-\frac{\partial\Gamma^{\mu}_{\nu\alpha}}
{\partial x^{\beta}}+\Gamma^{\mu}_{\gamma\alpha}\Gamma^{\gamma}_{\nu\beta}
-\Gamma^{\mu}_{\gamma\beta}\Gamma^{\gamma}_{\nu\alpha}
\label{eqn:riem}
\end{equation}
the nonzero components are found to be
\begin{equation} 
R^i_{0j0}=-R^i_{00j}=\frac{q}{mc^2}\left(\frac{\partial^2\Phi_E}{
\partial x^i\partial x^j}+\frac{1}{c}\frac{\partial^2A^i}{\partial t
\partial x^j}
-\frac{dx^n}{dt}\frac{\partial^2 A^n}{\partial x^i\partial x^j}
+\frac{dx^n}{dt}\frac{\partial^2 A^n}{\partial x^n\partial x^j}\right).
\label{eqn:Ri0j0}
\end{equation}
The only nonzero components of the Ricci curvature tensor
\begin{equation} 
R_{\mu\nu}=R^{\alpha}_{\mu\alpha\nu}
\label{eqn:ricci}
\end{equation}
is found to be
\begin{equation} 
R_{00}=\frac{q}{mc^2}\left[\nabla^2\Phi_E+\frac{1}{c}\frac{\partial}
{\partial t}\left(\vec\nabla .\vec A\right)
-\vec v.\nabla^2\vec A+\vec v.\vec \nabla \left(\vec\nabla .\vec A\right)\right],
\label{eqn:R00-1}
\end{equation}
where $\vec v$ is the velocity of the test particle. Using the equations
\begin{equation} 
\nabla^2\Phi_E+\frac{1}{c}\frac{\partial}{\partial t}
\left(\vec\nabla .\vec A\right)=-4\pi k_e\rho_Q,\hspace{0.5cm}\vec\nabla
\times\vec B=4\pi k_m\left(\vec J+\vec J_D\right),
\label{eqn:pot}
\end{equation}
where $k_e$ and $k_m$ are the electric (Coulomb) and the magnetic
(Biot-Savart) constants, $\rho_Q$ is the charge density, $\vec J$ is  the
ordinary current density, and $\vec J_D=1/(4\pi k_e)\partial\vec E/
\partial t$ is the displacement current density, Eq. (\ref{eqn:R00-1}) becomes
\begin{eqnarray} 
R_{00}&=&-\frac{q}{mc^2}4\pi \left[k_e\rho_Q-k_m\vec v.\left(
\vec J+\vec J_D\right)\right]\nonumber \\
 &=& -\frac{q}{mc^2}4\pi k_e\left[\rho_Q-\frac{1}{c^2}\vec v.\left(
\vec J+\vec J_D\right)\right],
\label{eqn:R00-2}
\end{eqnarray}
where $k_m/k_e=1/c^2$ has been used. Noting that
$R_{00}=\partial\Gamma^i_{00}/\partial x^i$ and
$\Gamma^i_{00}=(g^{ii}/2)(-\partial g_{00}/\partial x^i)$ it follows that
\begin{equation} 
R_{00}\approx -\frac{1}{2}\nabla^2g_{00},
\label{eqn:R00-3}
\end{equation}
where we have set $g^{11}=g^{22}=g^{33}\approx 1$. Assuming
$\nabla^2\vec v=0$ and $\vec\nabla .\vec A=0$, Eqs. (\ref{eqn:R00-2})
and (\ref{eqn:R00-3}) are satisfied by
\begin{equation} 
g_{00}\approx -\left(1+2\frac{q}{m}\frac{\Phi_E}{c^2}\right).
\label{eqn:g00-1}
\end{equation}
In a region where there are superimposed gravitational and electromagnetic fields, 
the above procedure would give
\begin{equation} 
ds^2=-\left(1+2\frac{\Phi_G}{c^2}+2\frac{q}{m}\frac{\Phi_E}{c^2}\right)c^2dt^2+2\frac{q}{m}\vec A.d\vec xdt...,
\label{eqn:ds-1}
\end{equation}
where $\Phi_G$ is the gravitational potential
\footnote{Note that the emergence of the term 
$(\vec A.d\vec x)dt$   is similar to
what happens in Gravitomagnetism \cite{mashhoon, clark} where there is a gravitomagnetic 
vector potential $\vec A_g$ like the electromagnetic vector potential $\vec A$.}
.\\

Lo and behold, these equations  reveal that a distribution of electric
charge curves the spacetime just like a neutral mass distribution does
(apart from the magnitude and a possible difference in the sense of the
curvature). The motion of a test charge in an electromagnetic field
is thus geometrized by connecting electromagnetic potentials to the
metric of the spacetime. One distinct feature different from gravitation
is that test particles with different $q/m$'s  have their
own geometries in the same electromagnetic field, whereas all test
particles have the same geometry in a gravitational field irrespective
of their masses. 

\section{A Charged Test Particle in the Viscinity of a Charged Sphere
\label{sec:sphere}}

Another argument which gives the same $g_{00}$ presented above is as follows.
 Consider an electrically charged metallic sphere of mass $M$ and charge $Q$. 
 Let there be a test charge $q$ of mass $m$ in the superimposed gravitational and
electrical potentials of the sphere. For simplicity, let us assume that the test
charge is at rest at a position $r$ from the center of the sphere. The energy of
the test charge is given, upon factoring $m_ic^2$ out, by
\begin{eqnarray} 
E&=&m_ic^2\left(1-\frac{m_g}{m_i}\frac{GM}{c^2r}-\frac{q}{m_i}\frac{k_eQ}{c^2r}\right)\nonumber \\
 &\approx& m_ic^2\left(1-2\frac{m_g}{m_i}\frac{GM}{c^2r}+2\frac{q}{m_i}\frac{k_eQ}{c^2r}\right)^{1/2},
\label{eqn:Energy}
\end{eqnarray}
where it has been assumed that the second and the third terms on the right are much 
smaller than one. Comparing this with the general relativistic exact equation for
$Q=0$ \cite{landau}
\begin{equation} 
E=\frac{mc^2(-g_{00})^{1/2}}{\sqrt{1-v^2/c^2}}
\end{equation}
we obtain the correct expression for $g_{00}$ upon putting $v=0$. There is no reason why this
purely classical argument, which is correct in the gravitational case, should fail when 
it is extended to include the electric potential energy of the test charge. Therefore we
obtain
\begin{equation} 
g_{00}= -\left(1-2\frac{GM}{c^2r}+2\frac{q}{m}\frac{k_eQ}{c^2r}\right),
\label{eqn:g00-E}
\end{equation}
where the ratio $m_g/m_i$ has been set to one.

\section{The Action Integral for a Charged Particle
\label{sec:action}}
Another supporting clue for the unified description of gravitation and
electromagnetism in the manner we contemplate comes from the action
integral for a charged particle moving in a region where there are
superimposed gravitational and electromagnetic fields. The relativistic Lagrangian for a test
particle of mass $m$ and electric charge $q$ is
\begin{equation} 
{\cal L}=-mc^2\sqrt{1-\frac{v^2}{c^2}}-m\Phi_G-q\Phi_E+q\vec A.\vec v
\label{eqn:lagr}
\end{equation}
where $\Phi_G$ and $\Phi_E$ are the gravitational and electrical
potentials. Even though there is no experiment to support it, the
prevailing assumption in the literature dictates the action
\begin{equation} 
I=\int^{t_2}_{t_1}{\cal L}dt
\label{eqn:act1}
\end{equation}
corresponding to ${\cal L}$ in eq.(\ref{eqn:lagr}) to be written as
\begin{equation} 
I=\int^{t_2}_{t_1}\left(-mc\frac{ds}{dt}-q\Phi_E+q\vec A.\vec v\right)dt,
\label{eqn:act2}
\end{equation}
where $-mcds/dt$ contains only the first two terms in Eq. (\ref{eqn:lagr})
\cite{landau}.
Emancipating ourselves from this assumption and including all the terms of
${\cal L}$ in $-mcds/dt$ entails
\begin{eqnarray} 
ds^2&=&-\left(1+2\frac{\Phi_G}{c^2}+2\frac{q}{m}\frac{\Phi_E}{c^2}-
2\frac{q}{m}\frac{\vec v.\vec A}{c^2}\right)c^2dt^2+...,\nonumber \\
&=&-\left(1+2\frac{\Phi_G}{c^2}+2\frac{q}{m}\frac{\Phi_E}{c^2}\right)c^2dt^2+
2\frac{q}{m}\vec A.d\vec xdt+....
\label{eqn:ds-2}
\end{eqnarray}
where terms that vanish as $c\rightarrow\infty$ have been dropped. The idea
then suggests itself that the $g_{00}$ component of the metric tensor
$g_{\mu\nu}$ is
\begin{equation} 
g_{00}=-\left(1+2\frac{\Phi_G}{c^2}+2\frac{q}{m}\frac{\Phi_E}{c^2}\right),
\label{eqn:g00-2}
\end{equation}
which indicates again that the electromagnetic field
curves the spacetime on the same footing as the gravitational field. In
the current GR theory the metric tensor $g_{\mu\nu}$ is also interpreted as
the gravitational field proper. Since the components of $g_{\mu\nu}$
are determined sufficiently by the Einstein field equations it is believed
that there is no room for the electromagnetic field
in the same geometry. Our treatment of the electromagnetic field
a la Cartan, the elevator experiments, and the $g_{00}$ we have obtained in
Eqs. (\ref{eqn:ds-1}), (\ref{eqn:g00-E}), and (\ref{eqn:g00-2}) indicate that this
interpretation may not be correct. We are motivated by  Eqs.(\ref{eqn:R00-3}) and (\ref{eqn:g00-1})
to suggest that they correspond to the Newtonian limit of
more fundamental tensor equations involving the Ricci tensor $R_{\mu\nu}$.
Before we write down these equations, to convince ourselves more let us
present the elevator cabin thought experiments by replacing the
gravitational field by an electromagnetic field. The situation is very much
like that in the hypothetical gravity with $m_g\neq m_i$.

\section{The Elevator Thought Experiment for Charged Particles
\label{sec:elevator}}
Inasmuch as there exists a local inertial frame for every particle or
particles with the same field charge-to-inertial mass ratio, we shall
consider only one test particle in the following. Consider again a closed
and stationary elevator cabin with an observer and a test particle in
it. Let the elevator, the observer, and the test particle have the same
electric charge-to-mass ratio $q/m$. For simplicity and definiteness
assume that all the
charges are positive. Let there be no gravitational field but an external
downward uniform electric field $\vec E$ act on the system. When released
from rest by the observer, the test particle will move downward with an
acceleration $a=(q/m)E$. Now, let this system be moved into space where
there are no fields of any kind to act on it, and let it be accelerated
upward
\footnote{The rule for the direction of the acceleration of the cabin
is that it be opposite the direction of motion of the test particle.}
by an external agent with an acceleration whose magnitude is equal to that
above. The floor of the elevator will accelerate towards the test
particle released by the observer from rest. From the point of view of the
observer the static elevator and the accelerated elevator situations are
equivalent. Under these conditions he/she cannot distinguish between the
existence of the electric field and the acceleration of the elevator. The
single-particle equivalence principle for the electric field may thus be
stated as: {\em it is impossible to distinguish the fictitious inertial
forces from the real electric forces in a local region containing a single
particle or particles with the same electric charge-to-mass ratio.}   We are
proposing this equivalence principle because it seems to correspond to reality. 
After all, when a collection of particles with different $q/m$'s are released from rest
in a uniform electric field they will form groups as they accelerate
according to their $q/m$'s. Each such group of particles will have the
same spacetime manifold and thus may be taken collectively as test
particles
\footnote{It is assumed, as usual, that the interactions between the
particles are  negligible.}.  Let us also note that in the case when the
elevator is let to fall freely in the downward electric field considered
above, the test particle released will float as if the elevator were
motionless in free space. Again, since the acceleration of the test
particle relative to the elevator (or the observer) has ceased because
\begin{equation} 
a_{rel}=\left(\frac{q}{m}-\frac{Q}{M}\right)E=0,
\label{eqn:elev2}
\end{equation}
where $Q$ and $M$ are the electric charge and the mass of the elevator,
the elevator constitutes a local inertial frame. 

Next, let us consider the same elevator and its contents in a region where
there is only a uniform downward magnetic field $\vec B$. Let the test
particle be released horizontally with velocity $v$ towards the front
wall of the elevator. As the observer faces the front wall, he/she will
see the particle deflect counterclockwise towards his/her left with an
acceleration $a=|(q/m)\vec v\times \vec B|=(q/m)vB$. Afterwards, let the 
elevator be rotated uniformly in a clockwise fashion by an external agent 
(e.g. a rigid rod of length $R$ attached to the top of the elevator, the other 
end of the rod being fixed and serving as the rotation center.) with angular 
velocity $\vec\Omega$ which causes a Coriolis acceleration equal 
to that above, i.e.  $a=|2\vec v\times \vec\Omega| = 2v\omega_L = (q/m)vB$, provided the 
angular frequency $|\vec \Omega|=\omega_L=(q/2m)B$ is small enough so as to neglect the 
centrifugal acceleration $\omega^2_LR=(q/2m)^2B^2R$ compared to the Coriolis acceleration 
resulting in the condition $\omega_L\ll 2v/R$. The test particle when released from 
rest will be seen by the observer 
to be moving in exactly the same way as in the first situation 
\footnote{This argument has been further developed by Mashhoon into the gravitational
analog of Larmor's theorem \cite{mash}}.
Next, let the elevator, the observer, and
the test particle, all having the same electric charge-to-mass ratio, be
set into motion with identical velocities perpendicular to the downward magnetic
field. The elevator and its contents will move in circles of the same
radius but with different centers. The observer will see the test particle
float and hence the elevator constitutes a local inertial frame because
\begin{equation} 
a_{rel}=\left(\frac{q}{m}-\frac{Q}{M}\right)vB=0.
\label{eqn:elev3}
\end{equation}
We can now postulate the equivalence principle for the electromagnetic
field: {\em All effects of a uniform electromagnetic field locally on a
single particle or particles with the same electric charge-to-mass ratio
are identical to the effects of a uniform acceleration of the reference
frame.} The proposed extension of the equivalence principle to electromagnetic fields as presented 
here is actually already well supported by an important set of experimental evidences 
such as the Witteborn - Fairbank experiment to determine the acceleration of electrons
in a vacuum enclosed by a copper tube in the gravitational field of the earth and the 
London moment in rotating superconductors, discussed in \cite{matos}.

\section{The Gravitational and Electromagnetic Field Equations
\label{sec:gravitational}}
A very simple and experimentally testable unified description
of gravitation with electromagnetism may be rendered possible if we give
up the interpretation that $g_{\mu\nu}$ is the gravitational field
proper. We should accept the fact that $g_{\mu\nu}$ is simply the metric
tensor, to which gravitational as well as elecromagnetic fields contribute
separately but similarly, through which the spacetime curvature is
determined. Accepting this interpretation, we can immediately write down
the modified field equations. Consider a compact object with a distribution
of total mass $M_o$
and charge $Q_o$. Let there be a distribution of charged matter with total
mass $M$ and charge $Q$ {\em outside} this compact object. Let also a test
particle of mass $m$ and charge $q$ be moving in this region. The modified
field equations outside the compact object are
\begin{equation} 
R^{\mu\nu}-\frac{1}{2}g^{\mu\nu}R=\frac{8\pi G}{c^4}T^{\mu\nu}_M+
\frac{8\pi k_e}{c^4}\frac{q}{m}T^{\mu\nu}_{CC},
\label{eqn:modif}
\end{equation}
which should be compared with the Einstein's equations for this case
\begin{equation} 
R^{\mu\nu}-\frac{1}{2}g^{\mu\nu}R=\frac{8\pi G}{c^4}\left[T^{\mu\nu}_M+
T^{\mu\nu}_{EM}(Q)+ T^{\mu\nu}_{EM}(Q_o)\right].
\label{eqn:ein-2}
\end{equation}
 Here $T^{\mu\nu}_M$
is the matter  energy-momentum tensor
\footnote{We emphasize that {\em mass} has the same meaning here as in
Einstein's GR. Gravitational, electromagnetic, nuclear, and other forms
of energy may contribute to mass. Massless particles like photons have
an {\em effective mass} in a gravitational field given by $E/c^2$, with
$E$ being the energy of individual photons.}
 of the matter outside the compact object. $T^{\mu\nu}_{EM}(Q)$ and
$T^{\mu\nu}_{EM}(Q_o)$ are the
energy-momentum tensors of the electromagnetic fields due to $Q$ and
$Q_o$, respectively.
$T^{\mu\nu}_{CC}$ is a tensor
such that, in the absence of the  object and neglecting gravity,
$R^{00}_{CC}=(8\pi k_e/c^4)q/m\left(T^{00}_{CC}+T_{CC}/2\right)$
reduces to Eq. (\ref{eqn:R00-2}) in the
Newtonian limit, with $R^{00}_{CC}$ being the contribution of the second
term on the right in Eq. (\ref{eqn:modif}) to $R^{00}$, and   $T_{CC}$
being the
trace of $T^{\mu\nu}_{CC}$. The tensor $T^{\mu\nu}_{CC}$ may be called
the {\em charge-current tensor}. It replaces $T^{\mu\nu}_{EM}(Q)$ in
Eq. (\ref{eqn:ein-2})
with a different coupling. It is the source of the electromagnetic field
outside the object and does not contribute to the gravitational field.
One such tensor is
\begin{equation} 
T^{\mu\nu}_{CC}=\frac{1}{3}v_{\alpha}{\cal J}^{\alpha}
\left(\frac{1}{c^2}U^{\mu}U^{\nu}+g^{\mu\nu}\right),
\label{eqn:charget}
\end{equation}
where $v^{\alpha}=\left(\gamma_vc, \gamma_v\vec v\right)$ is the
four-velocity of the test particle,
 ${\cal J}^{\alpha}=\left(c\rho_{Q}, \vec J+\vec J_D
\right)$, and $U^{\mu}=\left(\gamma_uc, \gamma_u\vec u\right)$ is the
four-velocity of the charge distribution, and
$\gamma_{v(u)}=\left(1-v^2(u^2)/c^2\right)^{-1/2}$.
It is clear that $T^{\mu\nu}_{CC}$ should not be confused with an
energy-momentum tensor.
In the Newtonian limit, when $v/c<<1$, $T^{00}_{CC}=0$.
Notice that the second term on the right - hand side of Eq. (\ref{eqn:ein-2}) is replaced by the 
$T_{CC}^{\mu\nu}$ term in Eq. (\ref{eqn:modif}) while the   third term on the right - hand side of 
Eq. (\ref{eqn:ein-2}) which is due to the
electromagnetic field of the  object does not appear in our scheme. This is
due to the same reason that the energy-momentum tensor of the gravitational
field of the  compact object does not appear on the right - hand sides of the
Eqs. (\ref{eqn:modif}) and (\ref{eqn:ein-2}).
In our scheme the effects of the free gravitational and electromagnetic
fields (which are due to the compact object) on the
spacetime geometry are already implicitly included by the left-hand side
of the equations.
To avoid any confusion, let us emphasize once again that the
first term on the right-hand side of Eq. (\ref{eqn:modif}) is the matter
tensor due to mass-energy and is the only source of gravity. The second term,
with a different coupling from the first, is the charge-current tensor and
contributes to the electromagnetic field only. This is because, in our scheme,
the unified GR described
by Eq. (\ref{eqn:modif}) is the metric theory of gravitation and
electromagnetism. Of course,
there is the possibility that it might also be the metric theory of weak and
strong interactions.
In the presence of weak and strong charges the right-hand
side of Eq. (\ref{eqn:modif}) may actually be
containing terms whose forms we do not know at  present. However, for
this to happen the form of these interactions must be similar to those
of gravity and electromagnetism. Currently, this seems not to be the case
and is an open question that deserves further study
\footnote{It seems that  all the interactions related to forces representable by potentials
curve the spacetime independently.}.

We emphasize that the trajectory of a charged particle moving in  superimposed
gravitational and electromagnetic fields due to a given source only is not described, in our scheme, by
the equation
\begin{equation} 
\frac{d^2x^{\mu}}{ds^2}+\Gamma^{\mu}_{\alpha\beta}\frac{dx^{\alpha}}
{ds}\frac{dx^{\beta}}{ds}=\frac{q}{mc^2}F^{\mu}_{\alpha}
\frac{dx^{\alpha}}{ds},
\label{eqn:geo-G-EM}
\end{equation}
where $F^{\mu\alpha}$ is the electromagnetic field strength tensor. The
correct equation for the trajectory of such a particle is the geodesic
equation (\ref{eqn:geo}) in which, contrary to Einstein's GR, the
coefficients $\Gamma^{\mu}_{\alpha\beta}$ get direct contribution from the
electromagnetic field. In Einstein's GR, in the absence of gravity or in
the presence of gravity but locally, the equation of motion of a charged
test particle in an electromagnetic field, due to the vanishing of the
$\Gamma^{\mu}_{\alpha\beta}$, is the special relativistic equation
\begin{equation} 
\frac{d^2x^{\mu}}{ds^2}=\frac{q}{mc^2}F^{\mu}_{\alpha}\frac{dx^{\alpha}}
{ds}.
\label{eqn:geo-SR}
\end{equation}
In our scheme, the equation of motion of a charged particle in such a case
is
\begin{equation} 
\frac{d^2x^{\mu}}{ds^2}+\Gamma^{\mu}_{\alpha\beta}\frac{dx^{\alpha}}
{ds}\frac{dx^{\beta}}{ds}=0,
\label{eqn:geo-UGR}
\end{equation}
which is the geodesic equation (\ref{eqn:geo}). But now the coefficients
$\Gamma^{\mu}_{\alpha\beta}$ get contribution from the electromagnetic
field only. Hence, in our scheme Eq. (\ref{eqn:geo-SR}) is not exact
but approximate.

\section{The Line Element for a Spherically Symmetric Distribution of
Matter and Charge
\label{sec:line}}

To gain further insight into our scheme, let us consider
the field equations describing the empty space
\footnote{{\em Empty space} in our scheme means that there is neither
neutral nor charged matter present and no physical fields except the
gravitational and electromagnetic fields.}
external to a distribution of total mass $M$ and charge $Q$ (the subscript
$o$ has been dropped now). They are
\begin{equation} 
R^{\mu\nu}=0,
\label{eqn:empty}
\end{equation}
in our scheme, as opposed to
\begin{equation} 
R^{\mu\nu}=\frac{8\pi G}{c^4}T^{\mu\nu}_{EM}(Q),
\label{eqn:empty-charge}
\end{equation}
which are the Einstein field equations in this case.
In Newtonian gravity and electromagnetism the equations satisfied by the
potentials $\phi_G$ and $\phi_E$ in a region where there are
gravitational and electromagnetic fields but devoid of matter and
electric charge are $\nabla^2\phi_G=0$ and $\nabla^2\phi_E=0$. Of the
two general relativistic equations, Eqs. (\ref{eqn:empty}) and
(\ref{eqn:empty-charge}), it is the former one that upholds both these
Newtonian equations.
To find the solution of Eq. (\ref{eqn:empty}) for  a static and spherical
distribution of matter of mass $M$ and  electric charge  $Q$ with a test 
particle of mass $m$ and electric charge $q$ in the viscinity,
we write the spacetime metric in the standard form (see, for example \cite{foster})
\begin{equation} 
ds^2=-A(r)c^2dt^2+B(r)dr^2+r^2d\theta^2+r^2sin^2\theta d\phi^2,
\end{equation}
where $A(r)$ and $B(r)$ are functions of $r$ to be obtained by solving the field 
equations, Eq. (\ref{eqn:empty}). Following the same
steps given in textbooks, the metric is obtained to be
\begin{equation}
ds^2=-\left(1+\frac{k}{r}\right)c^2dt^2+\left(1+\frac{k}{r}\right)^{-1}dr^2+r^2d\theta^2+r^2sin^2\theta d\phi^2,
\end{equation}
where $k$ is an integration constant to be determined. Far from the source, $r$ is 
approximately the radial distance and $-g_{00}=1+h_{00}$.
In this region $h_{00}$ is small and equals $k/r$. Since there are superimposed 
Newtonian potentials given by $\Phi=(m_g/m_i)\Phi_G+(q/m_i)\Phi_E=\Phi_G+(q/m)\Phi_E$ 
outside the source,
equations (\ref{eqn:ds-1}) and (\ref{eqn:ds-2}) suggest that $h_{00}=2\Phi/c^2$ 
and $k=-2GM/c^2+2(q/m)k_eQ/c^2$. The solution is thus found to be
\footnote{Note that, if desired, the factor $q/m$ may be set to +1 or
-1 by choosing the units appropriately. For example, we may choose for
electrons $e=m_e$, where $e=1.6022\times 10^{-19}C$ and
$m_e=9.1095\times 10^{-31}kg$, which gives $1C=5.6856\times 10^{-12}kg$.
This makes $q_e/m_e=-1$ at the expense of changing
$k_e=8.9875\times 10^9Nm^2C^{-2}$ to
$k_e'=2.7803\times 10^{32}Nm^2kg^{-2}$, and $k_m=10^{-7}Ns^2C^{-2}$ to
$k_m'=3.0935\times 10^{15}Ns^2kg^{-2}$. If gravitation and electromagnetism
are described together as we contemplate here, such a system of units seems
to be more natural. Note also that one could have measured the mass in
terms of the electric charge. This would have given for electrons,
 $1kg=1.7588\times 10^{11}C$, and $k_e$,  $k_m$, and $G$ would have changed
to $k_e'=1.5807\times 10^{21}m^3s^{-2}C^{-1}$,
$k_m'=1.7588\times 10^4mC^{-1}$, and
$G'=3.7935\times 10^{-22}m^3s^{-2}C^{-1}$.}
\begin{eqnarray} 
ds^2=-\left(1-2\frac{GM}{c^2r}+2\frac{q}{m}\frac{k_eQ}{c^2r}\right)c^2dt^2
+\left(1-2\frac{GM}{c^2r}+2\frac{q}{m}\frac{k_eQ}{c^2r}\right)^{-1}dr^2
+r^2d\theta^2+r^2sin^2\theta d\phi^2.
\label{eqn:newmetric}
\end{eqnarray}
When $Q=0$, or $q=0$ (but $m\ne 0$) Eq. (\ref{eqn:newmetric}) reduces correctly to 
the Schwarzschild solution \footnote{For neutral massless particles like photons it 
might so happen that $q/m=1$.} \cite{schw}.
When $Q\neq 0$, and $q\ne 0$, Eq. (\ref{eqn:newmetric}) replaces the Reissner-Nordstr\"om
solution \cite{rei,nord} of Eq. (\ref{eqn:empty-charge})
which we believe does not describe the actual physics correctly.
It is given by\footnote{It should be noted, however, that the Reissner-Nordstr\"om
solution is for a neutral test particle. The trajectory of a charged test particle 
in this case is given by Eq. (\ref{eqn:geo-G-EM}).}
\begin{eqnarray} 
ds^2=-\left(1-2\frac{GM}{c^2r}+\frac{Gk_eQ^2}{c^4r^2}\right)c^2dt^2+
\left(1-2\frac{GM}{c^2r}+\frac{Gk_eQ^2}{c^4r^2}\right)^{-1}dr^2+
r^2d\theta^2+r^2sin^2\theta d\phi^2.
\label{eqn:reis-nord}
\end{eqnarray}

Comparison of the third terms in  $g_{00}$ of Eqs. (\ref{eqn:newmetric})
and (\ref{eqn:reis-nord}) reveals
the philosophy of our unification. In Eq. (\ref{eqn:reis-nord}), the electric
field of the charge distribution contributes to the gravitational field of the matter.
Whereas in our scheme, which we call \emph{unified general relativity}
(hereafter UGR), there is an equivalence principle for the
electromagnetic field as well, and the right-hand side of
Eq. (\ref{eqn:empty}) is
zero, as opposed to Eq. (\ref{eqn:empty-charge}) of standard GR; the
electric field does not contribute to the gravitational field, it asserts itself separately.
One of the most salient features of the UGR theory is that it is a multi-metric theory. There
is a distinct metric for every test particle due to the charge-to-mass ratio, as there must be
so as to agree with classical electrodynamics in the Newtonian limit.
\section{Concluding Remarks
\label{sec:concluding}}
We have tried in this work a new route to geometrize the motion of a test
charge in an electromagnetic field. By considering the elevator cabin
experiments we were led to postulate an equivalence principle for the
electromagnetic field. Our geometrization was further supported by
the treatment of Newtonian electromagnetism in the language of curved
spacetime and by the action integral of a test charge. Within the precision
of the present experiments, the equivalence principle and the geometrization
of the motions of test particles in a given gravitational field are
{\em universal}. All test particles are effected universally. The
equivalence principle and the geometrization of the motion of test charges
in a given electromagnetic field, however, are {\em specific}. Different
test charges are effected specifically. This is, of course, what is
observed in nature.

Stipulating that the components of the metric tensor $g_{\mu\nu}$
correspond to the gravitational and electromagnetic potentials in
Riemannian spacetime, as opposed to, for example, Weyl's generalization
of Riemannian spacetime, we were led to give up the interpretation that
the $g_{\mu\nu}$ is the gravitational field only. This enabled us to postulate
the field equations describing the motion of a charged test particle in
superimposed gravitational and elecromagnetic fields. The new feature of
the modified field equations is the presence of what we call the
{\em charge-current tensor} $T^{\mu\nu}_{CC}$. This term on the right-hand
side of the field equations is the source of the electromagnetic field.
It is not an energy-momentum tensor and does not contribute to the
gravitational field. The expression we have presented for $T^{\mu\nu}_{CC}$
in this work has the correct Newtonian limit.
For a compact massive and charged object the space
outside it is not empty according to Einstein's GR; there is the
electromagnetic field contributing to the gravitational field through
its energy-momentum tensor. Whereas in our scheme, the space outside such
an object is empty because now there is an equivalence principle not only
for the gravitational field but also for the electric field. Two implications 
of the Principle of Equivalence for electromagnetic fields are (1) the Principle 
of General Covariance \cite{wein} is extended to include electromagnetic fields and (2) the
laws of Special Relativity hold locally in a coordinate system with vanishing 
gravitational and electromagnetic fields.

Physics is an experimental science. It is incumbent on a new theory that
possesses unorthodox predictions that it be confronted with experiment.
Unlike some other unsuccessfull unification schemes that had no new
experimental predictions, our scheme has several distinct predictions different
from that of standard GR. The deflection of an electron beam  in
the viscinity of a charged spherical mass in a vacuum chamber can be decisive 
to choose the correct line element among from the one we have proposed in this work
and the Reissner-Nordstr\"om line element which reduces to the Minkowski line element
for a laboratory - size charged sphere. An immediate prediction of the unified theory 
introduced is that there should be {\em electrical geometry waves} similar to 
{\em gravitational geometry waves} also known as {\em gravitational waves}.
There are other predictions of the present theory. For example, if one
considers an electron
moving radially away from a positively charged sphere and applies the
conservation of energy and then replaces the escape velocity from a radius
$r$ with the speed of light, one obtains the radius of the object from
which electrons cannot escape. Such an object can be rightly called
an {\em electrical black hole} and can be built in the laboratory. The
general relativistic theory to predict and give the radius of such an
object turns out to be the present theory. This and  other predictions
of the present theory will be presented in our forthcoming publications.\\

\noindent{\bf Acknowledgements}\\
\vspace{0.2cm}
We are grateful to Prof. Mahjoob O. Taha for invaluable 
discussions\footnote{ceased in the year 2000.}. Suggestions by Clovis Jacinto de Matos 
of ESA, Paris to improve the presentation are greatly appreciated.

\end{document}